\documentclass[nofootinbib,aps]{revtex4}
\usepackage{amssymb,amsmath,graphicx,color,microtype,bm}
\usepackage{graphicx}
\usepackage{longtable}
\usepackage{color,xcolor}
\usepackage{indentfirst}
\usepackage[utf8]{inputenc}

\newcommand{\be}{\begin{equation}}
\newcommand{\ee}{\end{equation}}
\newcommand{\beqn}{\begin{eqnarray}}
\newcommand{\eeqn}{\end{eqnarray}}
\DeclareMathOperator{\Real}{Re}

\usepackage[colorlinks=true, filecolor=red,citecolor=blue ]{hyperref}

\begin{document}
\title{Massive Fields as Systematics for Single Field Inflation}
\author{Hongliang Jiang}
\email{hjiangag@connect.ust.hk}

\author{Yi Wang}
\email{phyw@ust.hk}

\affiliation{Department of Physics, The Hong Kong University of Science and Technology, \\
  Clear Water Bay, Kowloon, Hong Kong, P.R.China}
\affiliation{Jockey Club Institute for Advanced Study, The Hong Kong University of Science and Technology, \\
  Clear Water Bay, Kowloon, Hong Kong, P.R.China}

\begin{abstract}
During inflation, massive fields can contribute to the power spectrum of curvature perturbation via a dimension-5 operator. This contribution can be considered as a bias for the program of using $n_s$ and $r$ to select inflation models. Even the dimension-5 operator is suppressed by $\Lambda = M_p$, there is still a significant shift on the $n_s$-$r$ diagram if the massive fields have $m\sim H$. On the other hand, if the heavy degree of freedom appears only at the same energy scale as the suppression scale of the dimension-5 operator, then significant shift on the $n_s$-$r$ diagram takes place at $m=\Lambda \sim 70H$, which is around the inflationary time-translation symmetry  breaking scale. Hence, the systematics from massive fields pose a greater challenge for future high precision experiments for inflationary model selection. This result can be thought of as the impact of UV sensitivity to inflationary observables.
\end{abstract}

\maketitle

\section{Introduction}

Inflation has been the leading paradigm for the early universe cosmology. Over the decades since inflation has been proposed, numerous inflation models have been developed (see \cite{Chen:2010xka, Wang:2013eqj} for reviews). It has been hoped that one can pin down the preferred inflation model among the a few major ones, by precision tests of $n_s$ and $r$.

Unfortunately, there is a rich variety of possible physics which can happen during inflation and change the prediction of simplest inflation models. Those changes includes:
\begin{itemize}
    \item Reheating. This is the best discussed systematics so far. Depending on the reheating time (in e-folds after the end of inflation), reheating temperature, and the geometry of reheating surface in field space \cite{Sasaki:2008uc, Huang:2009vk} and position space \cite{Dvali:2003em,Kofman:2003nx}, the predictions of single field inflation can change by large amount.
    \item Not-so-observable e-folds. The first 10 e-folds of observable inflation has been very well obseved by CMB experiments. But at much smaller scales, closer to reheating, the physics is much harder to extract. There are much fewer and looser constraints on what have happened (such as primordial black holes \cite{Hawking:1974rv}). However, note that prediction of inflation models on the $n_s$-$r$ diagram does not only depend on the observable e-folds, but also depend on the total e-fold of inflation, and thus the no-drama assumption of the not-so-observable e-folds. For example, there can be change of potential, particle production \cite{Chung:1999ve, Flauger:2016idt}, multi-stream inflation \cite{Li:2009sp, Wang:2013vxa}, and so on. The new physics change the calculation of number of e-folds and thus change the prediction of inflation models on the $n_s$-$r$ diagram. Typically, the changes of this type are parametrically simlar to the change of inflationary e-folds (shifting the $n_s$-$r$ diagram in a similar way).
    \item Hidden sectors during inflation. For example, the impact on the $n_s$-$r$ diagram from non-Gaussian hidden sectors via the short-long modes coupling  has been discussed in \cite{Bonga:2015urq}. In  \cite{Tolley:2009fg,Achucarro:2012sm}, it is noted that the density fluctuations can become sensitive to the UV physics, due to integrating out massive fields and thus modifying the   sound speed. The impact on the $n_s - r $ diagram is studied in \cite{Achucarro:2015rfa}. The UV sensitivity of Higgs inflation from dimension-6 operators is discussed in \cite{Burgess:2014lza}.

\end{itemize}
All above systematics (except the uncertainty in number of e-folds, which has already been well studied) have non-trivial assumptions. There is no problem to compare the simplest inflation model with experiments, avoiding the additional assumptions.

However, there is one class of correction to the simplest inflation model, which can not be avoided. The fundamental theory of inflationary fluctuations is quantum field theory, which should be understood as an effective field theory~(EFT). At higher energy scales, additional operators appear. Those operators are at most suppressed by the Planck scale. In this paper, we consider the dimension-5 operator
\begin{align} \label{eq:o5}
  \mathcal{O}_5 = -\frac{1}{2\Lambda} (\partial\phi)^2  \sigma~, 
\end{align}
where $\phi$ is the inflaton field, and $ \sigma$ is the massive field with mass $m$. The mass  is naturally of order Hubble scale $H$.  This is the most likely mass range for addition fields due to standard model uplifting, symmetry breaking and non-minimal coupling. This setup of field content and operator is known as Quasi-Single Field (QSF) inflation \cite{Chen:2009we, Chen:2009zp, Baumann:2011nk}.   Note that even if the Hubble scale massive field is not there by assumption or due to some peculiar reason, naturally there should be some massive fields around the cutoff scale since it is  the  defining feature of EFT.

In this paper, we will focus on the systematics of massive field to the single field $n_s$-$r$ diagram, assuming null result of primordial non-Gaussianity (there is orders-of-magnitude room for future observation of non-Gaussianity. So this assumption actually show the importance for such a measurement). As the correction to the scalar power spectrum is calculated numerically in \cite{Chen:2009we, Chen:2009zp} and later analytically in \cite{Chen:2012ge}, this bias on the $n_s$-$r$ diagram is straightforward to calculate. Nevertheless, it is important to realise and study this systematics, to avoid misleading interpretation of the future precision measurements on $n_s$ and $r$.

The generality of this issue has been ignored in the literature. One of the reasons is that, one may think that the correction is likely to come from a much higher energy scale, and thus may be ignored. However, this is not true. Note that the correction of power spectrum from $\mathcal{O}_5$ comes from the two-point interaction
\begin{align}
\delta\mathcal{O}_5 = \frac{1}{\Lambda} \dot\phi_0 \delta\dot\phi \delta\sigma~. 
\end{align}
Here the background motion $\dot\phi_0$ is actually very significant. The second Friedmann equation tells that $\dot\phi_0^2 = 2 M_p^2 H^2\epsilon $. Thus even if we choose $\Lambda$ to be the scale of reduced Planck mass $M_p=\sqrt{\hbar c/(8\pi G)}=2.4\times 10^{18}$GeV, we still have
\begin{align}
\delta\mathcal{O}_5 = \sqrt{2\epsilon} H\delta\dot\phi\delta\sigma~,
\qquad \epsilon\equiv - \frac{\dot H}{H^2}~.
\end{align}
Thus, the correction of the inflationary power spectrum (which involves two such vertices in a perturbative calculation) is $\mathcal{O}(\epsilon)$ instead of the much stronger Planck-mass suppression. This is evident in the parameter choice of \cite{Chen:2009zp}, and further pointed out more explicitly in \cite{Assassi:2013gxa} for the impact of non-Gaussianity, but the impact on power spectrum (which is more dramatic in terms of shifting the prediction of single field inflation) was not emphasized.

Our goal  is to address and emphasize this issue  in this paper. As we will show, for  massive field with natural mass of order $H$, the correction is sizable  in $n_s-r$ diagram even if a Planck scale cutoff $\Lambda=M_p$ is assumed. At the  same time, even if one assumes that massive field does not appear until the breakdown of EFT, i.e. $m\gtrsim\mathcal O(\Lambda)$, a significant shift in $n_s-r$ is also considerable for $m=\Lambda \sim 70H$, slightly above the time-translation symmetry breaking scale of inflation. Therefore such corrections due to massive fields are not negligible, especially considering  the future high precision measurement with uncertainty $\delta n_s,\delta r \lesssim 0.001 $  \cite{Matsumura:2013aja,Mao:2008ug}.  Such corrections or shifts, if not properly identified,  source the systematic errors in the program of filtering inflation models with $n_s$-$r$ diagram and thus pose   a greater challenge in finding out the ``correct" inflation model. 

This paper is organized as follows. In Section \ref{sec:sri}, we set up the convention and review the phenomenology of $\phi^n$ and Starobinsky inflation. In Section \ref{sec:correction}, we compute the correction on the $n_s$-$r$ diagram from massive fields. We conclude in Section \ref{sec:conclusion}.
  
\section{Standard single field slow-roll inflation}
\label{sec:sri}

Inflation predicts a nearly scale invariant density perturbation 
\be
P_{\zeta}(k)=P_{\zeta}(k_*) \Big(\frac{k}{k_*} \Big)^{n_s-1}~,
\ee
where $k_*$ is the pivot scale. This has been confirmed by experiments and observationally, the  pivot power spectrum and spectral index are $P_{\zeta}(k_*)\approx 2.2\times 10^{-9}, n_s\approx0.96$ \cite{Ade:2015xua}.

Almost scale invariant gravitational waves are also generated during inflation
\be
P_{\gamma}(k)=P_{\gamma}(k_*) \Big(\frac{k}{k_*} \Big)^{n_t}, \qquad r=\frac{P_{\gamma}(k_*)}{P_{\zeta}(k_*)}~.
\ee
The search of primordial  gravitational waves will provide a  direct proof of inflation. Current experiments have put stringent constraint on the strength of primordial   gravitational  waves $r<0.07$ \cite{Array:2015xqh}.

A very broad class of interesting models for inflation is the single field inflation, which is theoretically simple and experimentally viable. In such a class of model, inflation is driven by the a scalar field known as inflaton with nearly flat potential. The slowly rolling of the inflaton gives rise to the exponential expansion of the universe with scale factor $a(t)\approx e^{Ht}$. The Hubble parameter $H$ is almost a constant during the quasi-de Sitter inflation period. 

The single-field slow-roll  inflation models are described by 
\be\label{SingInf}
S=\int d^4 x \sqrt{-g} \Big[  \frac{M_p^2}{2} R-\frac{1}{2}g^{\mu\nu} \partial_\mu \phi \partial_\nu\phi -V(\phi) \Big]~.
\ee 
This huge class of models  are parameterised by the slow-roll parameters
\be
\epsilon=-\frac{\dot H}{H^2}
=\frac{\dot \phi_0^2}{2M_p^2 H^2}
~, \qquad 
\eta=\frac{\dot\epsilon}{H\epsilon}~,
\ee
or in terms of the potential 
\be
\epsilon_V=\frac{M_p^2}{2}\Big( \frac{V'}{V}\Big)^2  ~, \qquad 
\eta_V=M_p^2 \frac{V''}{V}   ~.
\ee
They are related 
\be
\epsilon_V=\epsilon ~, \qquad
\eta_V=2\epsilon-\frac{1}{2} \eta~.
\ee 
In the slow-roll regime,   $\epsilon, |\eta|, \epsilon_V, |\eta_V|\ll 1$.

The single filed inflation models satisfy the following consistency relations:
  \beqn
   n_s-1&=&-2\epsilon-\eta=-6\epsilon_V+2\eta_V~,
\\
   r&=&16\epsilon= 16\epsilon_V~,
   \\
   n_t  &=& -r/8~.
   \eeqn

After inflation, the size of the universe has changed by a factor of $e^{H\Delta t}=e^N$ with e-folding number  
\be
N=\Big|\int_{\phi_i}^{\phi_f} \frac{1}{\sqrt{2\epsilon_V}}  \frac{d\phi}{M_p}\Big|=50 \sim  60~,
\ee
where initial field value $\phi_i$ is the related to the CMB, while $\phi_f$ is the field value at the end of inflation, essentially marked by  $\epsilon_V, |\eta_V| \sim 1$.
 
 In the next two subsections, we will review the very simple $\phi^n$ model and the experimentally favourable Starobinsky   model.   
   
\subsection{$\phi^n$ model}
 
For the simplest polynomial potential  $V(\phi)=A\phi^n$, one can calculate 
\be
\epsilon_V=\frac{n^2}{2} \Big(\frac{M_p}{\phi} \Big)^2 ~,\qquad
\eta_V=n(n-1)\Big(\frac{M_p}{\phi} \Big)^2 =\frac{2(n-1)}{n} \epsilon_V~,
\ee
   and
   \be
   N=\bigg|  \int_{\phi_i}^{\phi_f} \frac{\phi d\phi}{nM_p^2}\bigg|  
   =\bigg|  \frac{\phi^2}{2n M_p^2}\Big|_{\phi_i}^{\phi_f}\bigg|  
   =\bigg|  \frac{n}{4}\frac{1}{\epsilon_V}\Big|_{\phi_i}^{\phi_f}\bigg|  
   =\frac{n}{4}\frac{1}{\epsilon_V}-\frac{n}{4}\frac{1}{\epsilon_{V f}}~.
   \ee
   
   Thus, 
   \be
   \epsilon_V=\Big( \frac{1}{\epsilon_{Vf}}+\frac{4}{n}N \Big)^{-1}
   \ee
   where  $\epsilon_{V f}=1 $ marks the end of inflation. Using the consistency relation, one get
   \beqn
   n_s-1&=& -6\epsilon_V+2\eta_V
   =\frac{-2n-4}{n}  \epsilon_V~,
  \\
   r&=& 16\epsilon_V~,
   \eeqn
which can thus be shown in the $n_s-r$ diagram as a function of e-folding number $N$. See  the solid lines in Fig.~\ref{FigCase1} and Fig.~\ref{FigCase2}.  It is clear to see that the different models give different predictions in $n_s-r$ diagram. By comparing with the experimental data for $n_s$ and $r$, the program of filtering inflation models with $n_s-r$ diagram can be initiated \cite{Ade:2015lrj}.

  \subsection{Starobinsky $R^2$  inflation}
 The Starobinsky model was proposed in \cite{Starobinsky:1980te}   
 \be
 S=\frac{1}{2}\int d^4 x\sqrt{-g}  \Big(M_p^2 R+\frac{1}{6M^2}R^2  \Big)~.
 \ee

  Effectively, the Starobinsky   inflation can be reformulated in Einstein frame as an effective single-field inflation model described by Eq.~\eqref{SingInf} with potential \cite{Kehagias:2013mya}
  \be
  V(\phi)= \frac{3}{4} M_p^4 M^2 \Big[ 1-\exp\big( -\sqrt{\frac{2}{3}} \frac{\phi}{M_p}\big)\Big]^2~.
  \ee
  The slow-roll parameters, to leading order in $N$, are given by
\be
   \epsilon_V=\frac{3}{4N^2}~, \qquad
   \eta_V=-\frac{1}{N},
\ee   
and
\be
n_s-1=-\frac{2}{N}-\frac{9}{2N^2}~, \qquad
r=\frac{12}{N^2}~.
\ee
Thus, the Starobinsky model predicts a very small tensor-to-scalar ratio and is actually the most favourable model in the latest experiments \cite{Ade:2015lrj}.

\section{Single field inflation model interacting with massive field}
\label{sec:correction}

In spite of the simplicity and observational consistency of single field inflation, a single scalar degree of freedom can by no means be  the full picture of the early universe. At least the standard model particles should be there. Moreover, in the full-fledged quantum gravity like string theory, an infinite number of degrees of freedom are needed to make the whole theory consistent.  These degrees of freedom must play some role in the early universe, considering that that our universe originated from big-bang with extremely high temperature. So, single field inflation should be more appropriately   regarded as an effective field theory with one (scalar) degree of freedom.\footnote{Note that the EFT here is a little different from the EFT of \emph{quantum fluctuations} in \cite{Cheung:2007st}.  }
 The EFT is formally obtained by integrating out all the heavy massive fields. The effects of those massive fields can be recast  into the higher dimensional operators. 
 
 Instead of considering the EFT of single inflaton,  we will study the EFT of hidden massive particle and inflaton. This is to push  the cutoff to much higher energy scale, instead of the mass scale of lightest massive particle.

Suppose there is a hidden sector, which  interacts with inflator via a dimensional 5 operator \cite{Chen:2009we, Chen:2009zp, Assassi:2013gxa}
\beqn 
 S&=&\int d^4 x \sqrt{-g} \mathcal L~,\\
\mathcal L &=&\frac{M_p^2}{2} R +\mathcal L_{\phi}+\mathcal L_{ \sigma}+\mathcal L_{\text{mix}}[\phi, \sigma]~,\\
\mathcal L_{\phi} &=& -\frac{1}{2} (\partial \phi)^2-V(\phi)~, \\
\mathcal L_{ \sigma} &=&  -\frac{1}{2} (\partial\delta\sigma)^2-V_\sigma( \sigma)~, \\
 \mathcal L_{\text{mix}} &=& -\frac{1}{2\Lambda} (\partial \phi)^2 \sigma~.
\eeqn
 
The fluctuations on top of the classical de-Sitter background can be studied by decomposing the fields 
\be
\phi(t,\bm x)= \phi_0(t) +\delta \phi (t,\bm x) ~,\qquad \sigma(t,\bm x)= \sigma_0(t)+\delta\sigma(t, \bm x)~.
\ee

Then, the effective Lagrangian describing the fluctuations  is 
\beqn\label{LagEff}
\mathcal L_{\text{eff}}&=&-\frac{1}{2} (\partial \delta\phi)^2
-\frac{1}{2} (\partial \delta\sigma)^2-\frac{1}{2} m^2 \delta\sigma^2
+\rho \dot{\delta\phi}\delta\sigma+\cdots
\nonumber \\
&=&
-\epsilon M_p^2 (\partial\zeta)^2
-\frac{1}{2} (\partial \delta\sigma)^2-\frac{1}{2} m^2 \delta\sigma^2
-\sqrt{2\epsilon}M_p\rho \dot{\zeta}\delta\sigma+\cdots
\eeqn 
In the last line, we have translated the density fluctuations into the curvature fluctuations through
\be
\zeta(t,\bm x)=-\frac{H}{ \dot \phi_0} \delta\phi (t,\bm x)~,
\ee
and
\be
\rho=\frac{\dot\phi_0}{\Lambda}~.
\ee

In this paper, we adopt a very phenomenological approach and take  Eq.~\eqref{LagEff} as our starting point with \emph{almost free }parameters $m, \Lambda $, etc. 

The resulting  power spectrum $\bm P_\zeta$ (We use bold characters for quantities with massive field corrections) has been calculated in \cite{Chen:2009zp}
\beqn
\bm P_\zeta &=& P_\zeta \Big( 1+\delta \Big)  ~,
\qquad P_\zeta \equiv \frac{H^2}{8\pi^2 \epsilon M_p^2}~,
  \\    \label{deltaVal}
  \delta &\equiv&  2 \mathcal C \Big(\frac{\rho}{H} \Big)^2=4\mathcal C \Big( \frac{\sqrt{\epsilon}M_p}{\Lambda} \Big)^2
=
  \frac{1}{2\pi^2 P_\zeta} \mathcal{C} \Big( \frac{H}{\Lambda} \Big)^2~.
\eeqn
Note $P_\zeta $ is just the theoretical value of standard single field slow-roll inflation model, which becomes not a direct observable with the presence of massive fields. 
The analytical expression for the $\mathcal C$-function was obtained in \cite{Chen:2012ge}
\beqn
\mathcal C(m) &= &\mathcal C_1+\mathcal C_2 ~,\qquad \mu=\sqrt{\frac{m^2}{H^2}-\frac{9}{4}}~,\\
\mathcal C_1&=& \frac{\pi^2}{4\cosh^2(\pi \mu)}~, \\
\mathcal C_2&=&\Real\bigg\{ \frac{e^{\pi \mu}}{16 \sinh (\pi \mu)}  \Big[ \psi^{(1)}\Big( \frac{3}{4}+\frac{i\mu}{2}\Big) 
-\psi^{(1)}\Big(\frac{1}{4}+\frac{i\mu}{2}\Big)\Big]
-\frac{e^{-\pi \mu}}{16 \sinh (\pi \mu)}  \Big[ \psi^{(1)}\Big( \frac{3}{4}-\frac{i\mu}{2}\Big) 
-\psi^{(1)}\Big(\frac{1}{4}-\frac{i\mu}{2}\Big)\Big]
\bigg\}~.
\eeqn  
where  $\psi^{(1)}$ is the polygamma function 
\be
\psi^{(1)}(z)\equiv \frac{d^2 \Gamma(z)}{dz^2}~.
\ee
It is worth emphasising that the $\mathcal C$-function blows up and becomes singular in the small mass limit. In this case, we need to impose the e-folding number cutoff \cite{Chen:2009zp} and the correction of power spectrum scales as $N^2 \sim 3000$. This is a huge enhancement. We shall focus on the case of $m\gtrsim \mathcal O(H)$ thus without such enhancement, but one should keep in mind that a lighter mass can bring greater bias  (in this case the scenario looks more like multi-field inflation and a completely different prediction is indeed typical). 

Also, note that the $\mathcal C$-function is positive. Thus $\delta > 0$. As a result, the massive fields always enhance the theoretical scalar power spectrum: $P_\zeta < \bm P_\zeta$.

The massive fields modify the scalar sector while keep the fomula for tensor sector unaltered 
\footnote{It was argued in \cite{Kleban:2015daa} that the tensor power spectrum is very robust against all kinds of massive fields due to their rapid decay. Thus the inflationary energy scale  can be inferred from tensor power conclusively with the assumption of inflation and vacuum initial fluctuations (except if gravity is significantly modified \cite{Domenech:2017kno}). }. 
But note that to be consistent with the observed scalar power spectrum,  the massive field  contributions should be subtracted by lowering the height of original inflaton potential and thus the Hubble parameter. Correspondingly the value of the tensor power spectrum gets reduced. Thus, the ratio between scalar and tensor is indeed modified by a factor $  1+\delta  $ and has observational implications.  
 
Meanwhile, the spectral index also changes due to the presence of $\epsilon$-dependent correction $\delta$ 
\beqn
\bm n_s-1
\equiv
\frac{d \ln P_\zeta}{d \ln k}
  = -2\epsilon -\eta+  \eta \delta /(1+\delta)~.
 \eeqn

Thus, taking into account the effects of hidden massive fields, both the spectral index and tensor-to-scalar ratio in single field inflation model get renormalised  
 \beqn
 \bm n_s-1&=& -2\epsilon -\eta+  \eta \delta /(1+\delta)
 \approx -(6-4\delta) \epsilon_V+2(1-\delta)\eta_V~, \\
\bm r &=&\frac{16\epsilon}{1+\delta} \approx16\epsilon_V(1-\delta)~.
 \eeqn
Note that these two \emph{renormalized} parameters correspond to the observational values in experiments.  The impact is shown in Fig.~\ref{FigCase1} and Fig.~\ref{FigCase2} and will be elaborated in the followings.

\begin{figure}  
  \flushright
  \includegraphics[width=0.8\textwidth]{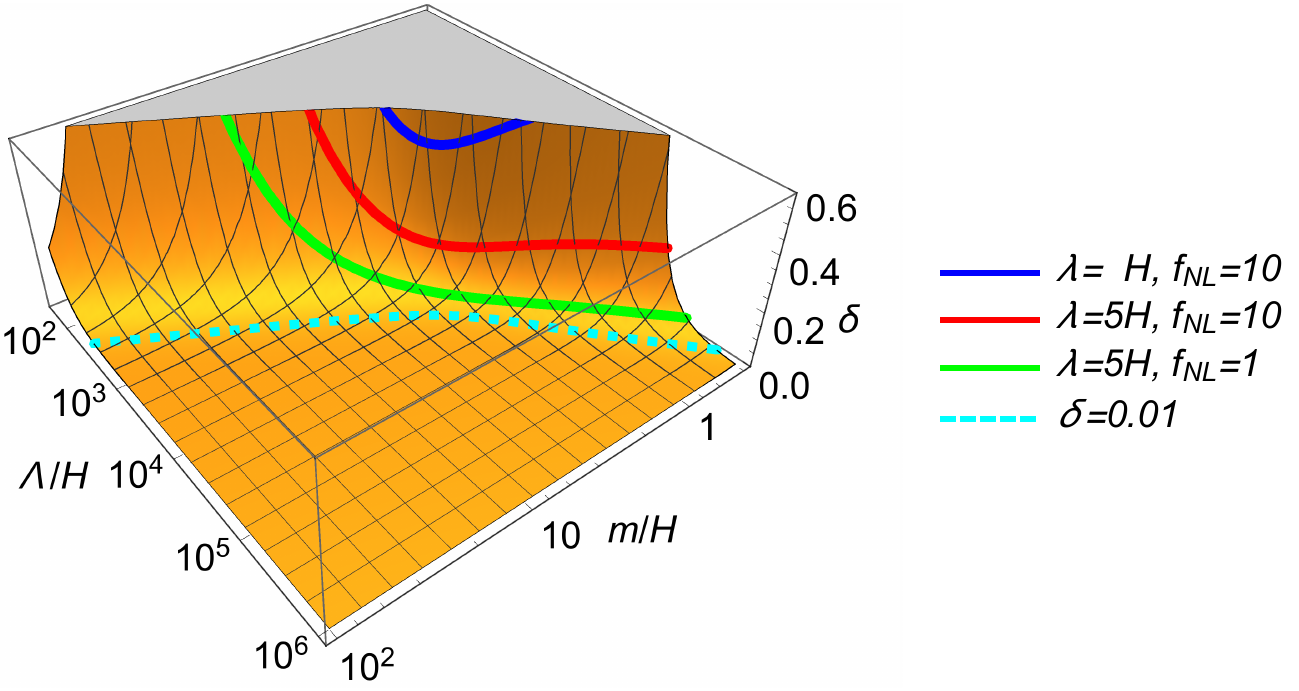}
  \caption{The dependence of $\delta$  on mass $m$ and cutoff $\Lambda$. Solid lines represent equal-$f_{\text{NL}}$ contours for specific self-coupling $\lambda$. Dashed line marks the threshold correction $\delta\sim 0.01$ for promising observations. Large-$\delta$ parts are chopped off since they are beyond the perturbative regime. 
    \label{Figdelta}}
  \end{figure}

Considering the future high precision experiments, the sensitivity on $ n_s$ and $ r$ can be as as high as $0.001$.  For example, the expected uncertainly $\delta  r \sim 0.001$ for the LiteBIRD \cite{Matsumura:2013aja}  and   $\delta   n_s<0.001$ through 21 cm tomography \cite{Mao:2008ug}. Thus, the corrections from massive fields indeed matter for these high precision experiments.   In Fig.~\ref{Figdelta}, we plot the correction $\delta$ as a function of mass $m$ and energy scale cutoff $\Lambda$. 

In the next two subsections, we will consider two interesting cases to illustrate the effects and show that the corrections are indeed significant even for some natural parameters.  

Before the discussion of different parameter regions of massive fields, let us review the impact of non-Gaussianity from massive fields  given current  tight experimental bounds. There are actually also cubic interactions in Eq.~\eqref{LagEff} which can source non-Gaussianities \cite{Chen:2009zp,Assassi:2013gxa}
\be\label{IntNonG}
\mathcal L_{\text{eff}} \supset \mathcal L_{\text{int}}= -\frac{(\partial \delta \phi)^2\delta\sigma}{2\Lambda}-\lambda( \delta\sigma)^3~.
\ee
The non-Gaussianity  corresponds to the first interaction term is $f_{\text{NL}}\sim \delta <1$ and thus can be ignored. For the second type interaction, the non-Gaussianity is of  equilateral shape, given by \cite{Chen:2009zp,Assassi:2013gxa,Gong:2013sma}
\beqn
f_{\text{NL}} &=&\frac{\mathcal{B}(m)}{2\pi} \frac{1}{\sqrt{P_{\zeta}}} \frac{\lambda}{H}  \Big( \frac{\rho}{H}  \Big)^3 \\
&\sim& \frac{240}{243} \frac{1}{16\pi \sqrt{P_\zeta}} \frac{\lambda}{H} \Big( \frac{\rho}{H} \Big)^3 \Big( \frac{H}{m}\Big)^6 
=\frac{30}{243} \frac{1}{ (2\pi)^4 P_\zeta^2} \frac{\lambda}{H} \Big( \frac{H}{\Lambda} \Big)^3 \Big( \frac{H}{m}\Big)^6 
\quad \text{   for } m \gg H~.
\eeqn
 In the large $m/H$ limit, the $\mathcal C$-function behaves like $H^2/(4m^2)$ \cite{Chen:2012ge}. Thus, Eq.~\eqref{deltaVal} reduces to
\be
\delta  \sim \frac{1}{8\pi^2 P_\zeta}\Big( \frac{H}{m} \Big)^2\Big( \frac{H}{\Lambda} \Big)^2~,
\ee
and then one gets
\be
f_{\text{NL}}\sim 10^{-9} \delta^3  \Big( \frac {\Lambda}{H} \Big)^3 \frac{\lambda}{H}~.
\ee
Note that although this relation  is derived  in the large mass limit, it should also be a good approximations for \emph{not-too-small mass} case. This is due the fact that essentially the mass dependent functions  $\mathcal{B}$ and $\mathcal C$ are the products of propagator of massive field. One propagator contributes $1/(m^2-p^2)\sim 1/m^2$. Since the Feynman diagrams corresponding to power spectrum correction and cubic interaction in Eq.~\eqref{IntNonG} consist of one and three massive field propagators respectively,   physically one  expects $\mathcal{B} \sim \mathcal C^3$  for \emph{a  wide range of mass}. It only breaks down for \emph{too-small-mass} whose  specific value is not useful in the current rough estimation.
 
Therefore, the small non-Gaussiaity does \emph{not} mean a small correction on the power spectrum. By assuming a not-too-high  energy scale cutoff or weak self-coupling, a significant observable shift in the $n_s-r$ diagram is possible even if the non-Gaussianity is small and undetectable. In Fig.~\ref{Figdelta}, experimentally promising parameter space  are shown and possible values of $\lambda$ are also suggested without violating the current experiment bound on non-Gaussianity $f_{\text{NL}}$.

\subsection{Case 1: $ m\sim H, \Lambda\sim M_p$}

This corresponds to the QSF inflation model proposed in \cite{Chen:2009zp}. Below the cutoff  $\Lambda$, there are two scalar degrees of freedom, inflaton and massive field. The hidden massive particles are produced through vacuum fluctuations, but then decay.   There are many models which can give massive fields with $m\sim H$, including turning trajectory \cite{Chen:2009zp}, supersymmetry breaking \cite{Baumann:2011nk}, compactification and non-minimal coupling.

QSF inflation predicts a unique family of shapes of non-Gaussianity \cite{Chen:2009we, Chen:2009zp, Baumann:2011nk, Assassi:2012zq, Kehagias:2015jha}, and such non-Gaussianity can be used to study the particle physics during inflation \cite{Arkani-Hamed:2015bza, Chen:2016nrs, Lee:2016vti, Chen:2016uwp, Chen:2016hrz}, and to probe the evolution history of the primordial universe \cite{Chen:2015lza, Chen:2016cbe, Chen:2016qce}. More observational aspects and other related topics can be found in \cite{Sefusatti:2012ye, Norena:2012yi, Assassi:2013gxa, Dimastrogiovanni:2015pla, Schmidt:2015xka, Meerburg:2016zdz} and \cite{Chen:2012ge, Noumi:2012vr, Emami:2013lma}.

\begin{figure} 
  \flushright
  \includegraphics[width=0.9\textwidth]{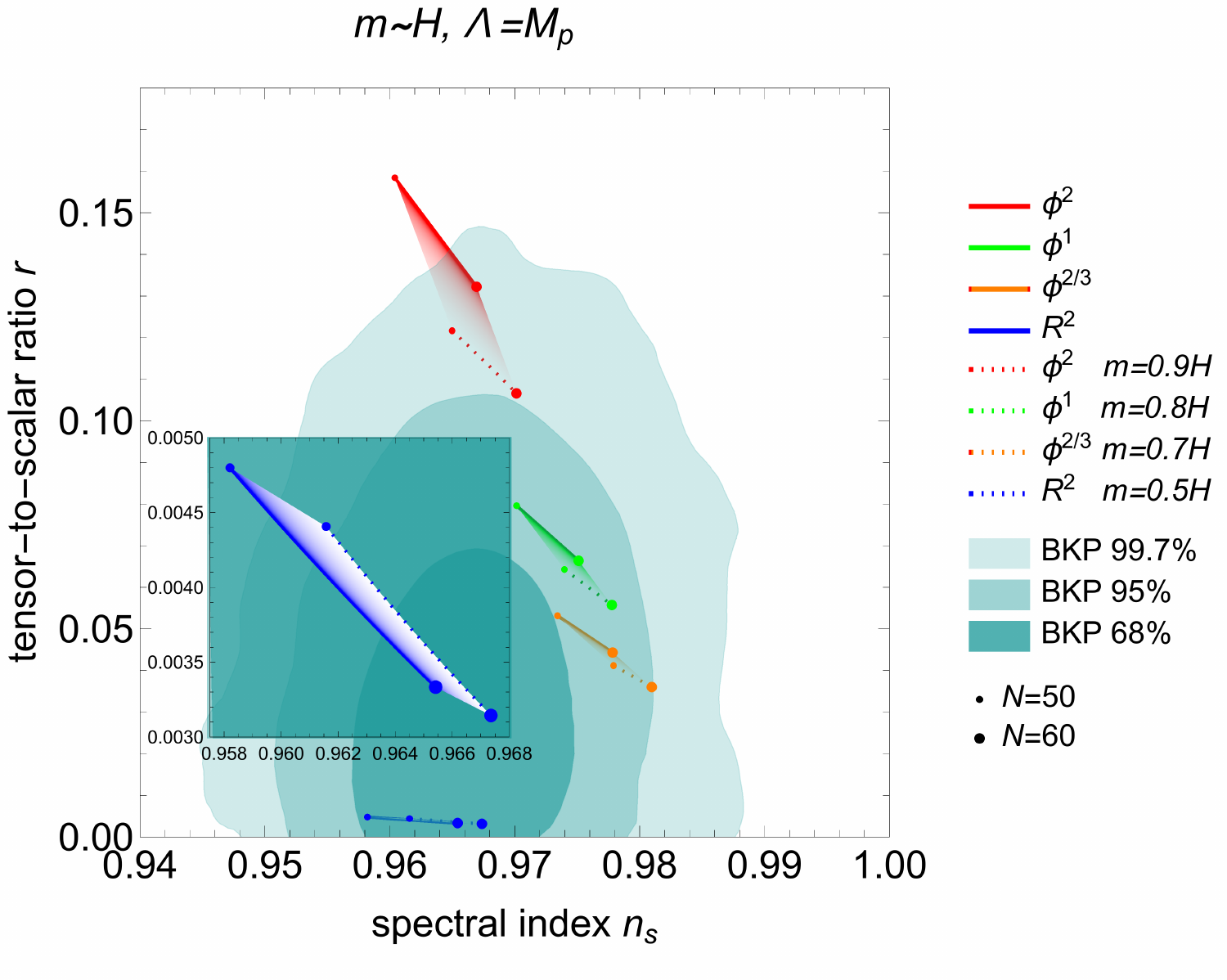}
  \caption{$n_s$ and $r$ for single field  inflation models, as well as their values after taking into account  the correction from massive fields with mass $m\sim H$ and cutoff $\Lambda=M_p$. The solid lines are the theoretical values predicted by  single filed inflation models. While the dotted lines comprise the massive field   correction of order $\delta\sim 0.2$ under specific choice of mass. The Starobinsky $R^2$ model is magnified in the subplot. The shaded contours at confidence level   68\%, 95\% and 99.7\% come  from BICEP2/Keck Array and Planck joint analysis which is the strongest experimental  bounds to date.
   \label{FigCase1}}
  \end{figure}

Since  our EFT is essentially an EFT with gravity,  the cutoff scale can be as high as $M_p$ in the ideal case. But beyond the Planck scale, the EFT breaks down and one needs to resort to a the full-fledged quantum gravity like string theory. 
Thus, we set $ \Lambda = M_p$ and illustrate the observational outcomes in $n_s-r$ diagram  Fig.~\ref{FigCase1}  for different models  discussed in Sec.~\ref{sec:sri}  and for  different choices of mass.

\subsection{Case 2: $ m=\Lambda < M_p$}

In this case,  there is only one scalar  degree of freedom below the cutoff scale. And the EFT becomes an EFT of single inflaton and one returns to the single field inflation paradigm. But the EFT  is  only valid below specific cutoff $\Lambda$. Above this cutoff and at high enough energy scale, the original EFT break down.  Near such energy scale, new degrees  of freedom kick  in and can not be integrated out anymore. The mass of these new degrees of freedom is expected to be of order cutoff scale $\mathcal O(\Lambda)$. From this perspective, we would like to consider the observational outcome for the case  $m=\Lambda$.

\begin{figure}
  \flushright
  \includegraphics[width=0.9\textwidth]{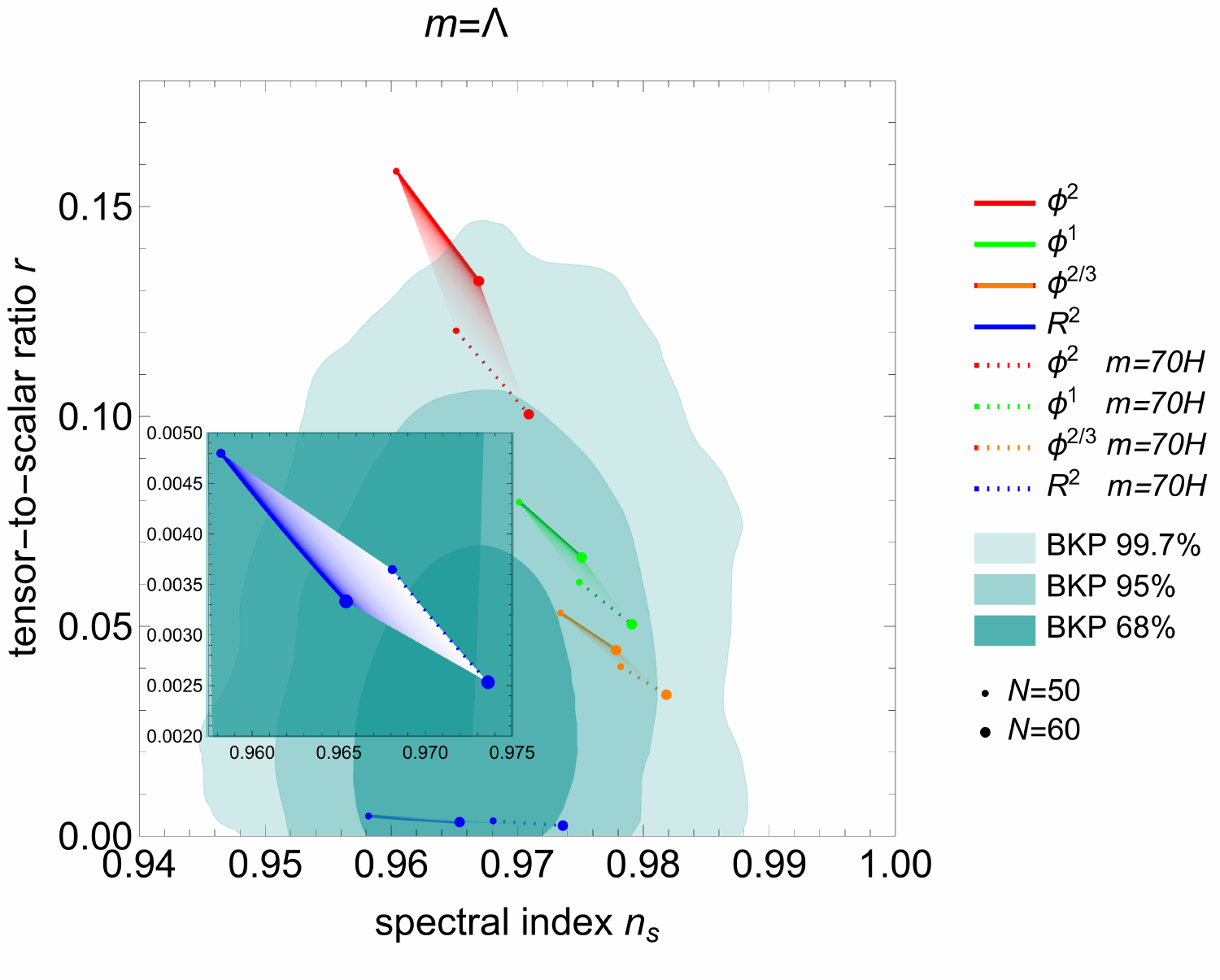}
  \caption{$n_s$ and $r$ for single field  inflation models, as well as their values after taking into account  the correction from massive fields with mass and cutoff $m=\Lambda=70H$. The dotted lines correspond to a correction of order $\delta\sim 0.24$. \label{FigCase2}}
    \end{figure}

In such a case,  from Eq.~\eqref{deltaVal} 
\be
\delta \sim \frac{1}{2\pi^2 P_\zeta} \mathcal{C} \Big( \frac{H}{m} \Big)^2~.
\ee
In the large $m/H$ limit, the $\mathcal C$-function behaves like $H^2/(4m^2)$ \cite{Chen:2012ge}. Since the correction from massive fields is expected to be very small in the perturbative regime, $P_\zeta $ is supposed  to be roughly the observational value $P_\zeta^{(\text{obs})}=2.2\times 10^{-9}$. Thus, 
\be
\delta \sim \Big(  \frac{49H}{m} \Big)^{4}~.
\ee
When $\Lambda=m\sim 70H$, the correction is of order $\delta\sim 0.24$. We present the corresponding corrections in the $n_s-r$ diagram in  Fig.~\ref{FigCase2}.

It should be noted that this scale $\Lambda \sim 70 H$ is around the symmetry breaking scale of inflation \cite{Cheung:2007st}
\be
\Lambda_b=|\dot\phi_0|^{\frac{1}{2}}\approx  60H~,
\ee
which is the  energy scale where the time-translation symmetry is spontaneously broken. Below such scale, an effective description of fluctuations in terms of Goldstone boson is applicable. Above such scale, other degrees of freedom can become relevant. The massive degree of freedom that we have considered indeed fits into this picture. Thus, although single field inflation is a good approximation or a very ``effective" EFT, some hints about the higher energy physics can still be inferred in experiments via the power spectrum. On the other hand, if not properly identified, such corrections pose a great challenge in future high precision experiment to select inflation model. (In this regime, one can actually integrate out the massive degree of freedom \cite{Tolley:2009fg,Achucarro:2010da,Achucarro:2012sm,Achucarro:2012yr,Gwyn:2012mw,Gwyn:2014doa} and the correction to the observables is discussed in \cite{Achucarro:2015rfa}.)

\section{Conclusion and Discussions}
\label{sec:conclusion}

To conclude, we have shown that for reasonably chosen parameters, massive fields significantly shift the position of a inflation model on the $n_s$-$r$ diagram. The impliciation  of such shift depends on the details of the inflation model. It may  save the model from current experiments; for example, from Fig.~\ref{FigCase1} and Fig.~\ref{FigCase2},  the $\phi^2$ model  deviates from the best fit by 3$\sigma$, but now with massive fields, it is saved to the boundary of the 2$\sigma$ contour. It may also make the model,  like   the $\phi^{2/3}$ model, less favourable as it used to be.
  
Due to such possible  meaningful shift, even the future precise measurement on $n_s$ and possible discovery of $r$ cannot precisely select between simple inflation models, if not considering the corrections from massive fields. Nevertheless, the uncertainty can be reduced if detections are made in some of the following experiments (even that, note that one point on the $n_s$-$r$ diagram correspond to a infinite number of inflationary potentials. Thus the $n_s$-$r$ diagram can only help to distinguish between some simple inflation models, instead of reconstruct the inflationary potential):

\begin{itemize}
  \item Non-Gaussianity. As the prediction of non-Gaussianity from QSF is unique, future search for non-Guassianity helps for eliminating the QSF systematics of single field inflation.
  \item Tilt of the tensor spectrum $n_t$. This is very difficult to measure, given that currently the primordial gravitational waves are not yet discovered. Nevertheless, if $n_t$ is measured, from the broken single-field consistency relation $n_t \neq -r/8$, one can in principle recover the contribution from QSF.
  \item Gravitational waves produced from reheating. This does not directly help to reduce the QSF systematics. However, better understanding of reheating, and thus e-folds of observable inflation, can reduce the e-fold uncertainty on the $n_s$-$r$ diagram and thus reduce the volume of uncertainties, which helps for comparing inflation model with data.
\end{itemize}
Note that these auxiliary experiments can also distinguish the massive field effects from other effects and thus enable us to tell whether the shift is caused by massive field or not. 

The correction due to the  hidden massive sector is supposed to be very general even  for other kinds of higher dimensional operator couplings, not limited to the  $\mathcal O_5$ in Eq.~\eqref{eq:o5} here. Such corrections are not completely negligible for some parameters even if the  Planck scale   cutoff is imposed as shown in Case 1. This just shows the nature of UV sensitivity of inflation.

In this work we have only considered the case where the massive field correction $\delta \ll 1$. Otherwise the perturbative calculation of massive field is not under-control. The corrections from large $\delta$ can be computed numerically using the equation of motion method \cite{Chen:2015dga}. In this case, the massive field contribution can dominate the inflationary power spectrum. The implication of $\delta>1$ can be interesting: some models which were already considered ruled out, such as $\lambda \phi^4$, may still fit the data. Also,  here we only consider one massive field for illustration purpose. However,  an infinite tower of massive degrees of freedom are expected in a UV complete theory of gravity, although there may be mass hierarchies.  And the total net observational outcome of these many massive fields is supposed to be the integrated effects which can obviously enhance  the corrections.  Finally, this work has assumed that the massive field correction appear at the tree level. This forbids the discussion of the impact of standard model particles during minimal non-Higgs inflation. This is because those standard model particles have to appear in loops due to charge conservation, assuming that the inflaton does not carry standard model charge. Also, more precise prediction from Higgs inflation can be computed, using the results in \cite{Chen:2016nrs, Chen:2016uwp, Chen:2016hrz}. We hope to address these issues in a future work.  

\bigskip
\noindent \textit{Acknowledgments.} This work was supported by the CRF Grants of the Government of the Hong Kong SAR under HKUST4/CRF/13G and ECS 26300316. YW would like to thank Fudan University and USTC for hospitality where part of this work was done.


\end{document}